# Phase Oscillation between Superfluid and Normal State of Neutrons in Neutron Stars — The Origin of Glitches of Pulsars[1]

Qiu-he Peng[a,b] ( qhpeng@nju.edu.cn )
Zhi–quan Luo [a,c]

[a] School of Physics and electronic information, China West Normal University,
   Nanchong, Sichuan, 637002, China
[b] Department of Astronomy, Nanjing University, Nanjing, 210093, China
[c] Department of Physics, Sichuan University, Chengdu, 610065, China

**Abstract**   Starting from a neutron star heating mechanism by the magnetic dipole radiation from the $^3P_2$ neutron superfluid vortices in neutron stars, we propose a neutron phase oscillation model which describes the phase transition between the normal neutron Fermi fluid and the $^3P_2$ neutron superfluid vortices at the transition temperature of $T_{trans} = (5-6) \times 10^8 K$. With this model, we can give qualitative explanation to most of the pulsar glitches observed up to date.

## 1. Introduction

It is shown from the observational results on the pulsar rotation period that some young pulsars experience a glitch (or macro jump) phenomenon. The regular pulse signals could be occasionally shorten by glitches at the typical amplitude of $\Delta\Omega_0/\Omega_0 \sim 10^{-6} - 10^{-10}$. These glitches are usually accompanied by a spin-down effect on the pulsar rotation period at a much larger rate: $\Delta\dot\Omega/\dot\Omega \sim 10^{-2} - 10^{-3}$ (Lyne et al. 2000). There are more than fifty glitches detected among twenty pulsars up to date. Eighteen of these glitches detected in eight of the glitch pulsars are great glitches with magnitude of $\Delta\Omega/\Omega > 10^{-6}$ (Lyne et al. 2000; Urama, 2002). Among these eighteen great glitches, nine of them have been detected from the PSR Vela in the last 26 years. There are also thirteen smaller glitches detected during last 23 years from the PSR Crab having smaller magnitude (Lin and Zhang, 2004). Besides these macro glitches, more micro-glitches with jump amplitude less than $10^{-12}$ have been detected in some pulsars.

[1] This research is supported by Chinese National Science Foundation No.10173005, No.10273006, and the Doctoral Program Foundation of State Education Commission of China

It is reported that there is a rough tendency that both the jump amplitudes of glitches and the frequency of glitches occur are decrease with increase of the pulse period when the pulsars become old (Lyne et al., 2000). No glitch pulsar with the period longer than 0.7 second has been detected.

A lot of research has been published on the origin of the pulsar glitches. Most proposed models are based on the interaction between the neutron (mainly $^1S_0$) superfluid vortices and the crust or the proton superconductor magnetic flux tubes (Baym,Pethick,and Pines, 1969; Anderson and Itoh, 1975; Alpar,Anderson,Pines and Shaham, 1981; Ruderman, Zhu and Chen,1998; Sedrakian and Cordes,1999). There is also a model based on the two stream superfluid instability (Andersson,Comer and Prix, 2003). However, the actual mechanism that triggers the glitches remains unspecified in virtually all these models (Andersson,Comer and Prix, 2003).

In this paper, we propose that the great glitches are triggered by a neutron phase oscillation between the normal Fermi fluid and the $^3P_2$ superfluid vortices at the phase transition temperature, $T_{trans} = (5-6) \times 10^8 K$. This phase oscillation is based on the our earlier research on the heating of neutron stars by the magnetic dipole radiation from the $^3P_2$ NSV (Peng, Huang and Huang,1980; Huang , Lingenfelter, Peng and Huang,1982).

## 2.  Neutron superfluid vortices (NSV)

### 2.1   Formation of neutron superfluid

The interior of a nascent neutron star (NS) is in chaos with highly turbulent classical vortices at temperature around $10^{11}K$. At the beginning of the nascent neutron stars, the neutron system is in a normal non-superfluid Fermi gas state. However, the interior temperature will rapidly decrease due to some physical cooling processes (Shapiro and Teukolsky, 1984). A phase transition from the normal Fermi gas to the $^1S_0$ neutron superfluid occurs when the temperature (T) drops below the corresponding critical temperature ($T_\lambda$ ($^1S_0$ (n))) of a phase transition (at the moment $t_0^{(S)}$)

$$T \leq T_\lambda(^1S_0(n)) = \Delta(^1S_0(n))/k \approx 2 \times 10^{10} K \qquad (1),$$

where $\Delta(^1S_0(n))$ is the neutron pairing energy gap in the $^1S_0$ state, $k$ is the Boltzmann constant . The density range of an isotropic $^1S_0$ neutron superfluid is $1 \times 10^{11} < \rho(g/cm^3) < 7 \times 10^{13}$. As the temperature decreases further, the an-isotropic $^3P_2$ neutron superfluid state appears (at the moment $t_0^{(P)} > t_0^{(S)}$) when the temperature decreases below another critical temperature ($T_\lambda$ ($^3P_2$ (n)))

$$T \leq T_\lambda(^3P_2(n)) = \Delta(^3P_2(n))/k \approx 6 \times 10^8 K \qquad (2)$$

In the density region where $7 \times 10^{13} < \rho(g/cm^3) < 1.4 \times 10^{14}$, another phase transition happens (Elgagøy, Engvik, Hjorth-Jensen and Osnes, 1996):

*Normal Fermi neutron system* □ $^3P_2$ *Superfluid*            (3)

### 2.2   Angular momentum deposited in NSV

It is widely recognized that angular moment can be hardly transferred out of a rapid collapsing core of a neutron star during the collapse period of a supernova explosion. It can be derived from the conservation law of angular momentum that a rotational collapsed core of supernova has an initial rotational period at sub-millisecond (Zheng et al. 2004). Besides, a significant fraction of the angular momentum of the rotational pre-supernova is reserved in the turbulent classical vortices of the nascent neutron star during supernova explosion. The classical vortices are immediately transformed to quantized (neutron) vortices once the superfluid appears in the neutron star.

The total angular momentum of the neutron star is

$$J_{tot} \sim 1 \times 10^{47} I_{45} (\Omega/10^2 s^{-1}) \qquad (4).$$

Where, $I_{45}$ is the moment of inertia of the neutron star in unit of $10^{45}$ g.cm², $\Omega$ is its rotating angular velocity.

The angular momentum of the crust is

$$J^{(Crust)} = M^{(Crust)} R_{NS}^2 \cdot \Omega \sim \overline{\rho}_{Crust} R_{NS}^4 \Delta R \cdot \Omega$$
$$\sim 1 \times 10^{39} (\overline{\rho}_{Crust}/10^8 g/cm^3) R_{NS,6}^5 (\Delta R/0.1 R_{NS})(\Omega/10^2 s^{-1}) \qquad (5).$$

Where, $R_{NS}$ is the radius of the neutron star. $R_{NS,6}$ is in unit of $10^6$ cm. $\Delta R$ is the thickness of the crust.

The interior of the neutron star is believed to rotate faster than the crust. The angular momentum of the anisotropic ($^3P_2$) superfluid revolving the axis of the NS is

$$J(^3P_2) \approx M(^3P_2) R^2(^3P_2) \cdot \Omega_{SF} \sim (4\pi/15)\overline{\rho}(^3P_2) R^5(^3P_2) \cdot \Omega_{SF}$$
$$\sim 2.3 \times 10^{41} \frac{\overline{\rho}(^3P_2)}{\rho_{nuc}} R_5^5(^3P_2)(\frac{\Omega_{SF}}{10^2 s^{-1}}) \qquad (c.g.s.) \qquad (6).$$

Where, $\Omega_{SF}$ is the rotating angular velocity of the superfluid around the rotation axis of neutron star. $\rho_{nuc}$ is the nuclear density in unit of $2.8 \times 10^{14}$ g/cm³. $R_5$ is the radius of the anisotropic superfluid in unit of $10^5$ cm.

The angular momentum deposited in all vortices (the neutron superfluid evolved around the axis of the vortex) is

$$J_{NSV} = N_{nuetron} \cdot \overline{n} h \approx 4.7 \times 10^{30} \overline{n} (\overline{\rho}(^3P_2)/\rho_{nuc}) R_{NS,6}^3 \qquad (c.g.s) \qquad (7).$$

Where, $n$ is the quantum number of superfluid vortex, $\overline{n}$ is its average value of all the corresponding vortices. $N_{neutron}$ is the total number in the superfluid state.

For the $^1S_0$ isotropic NSV, $\rho \sim 10^{12}$-$10^{13}$ g/cm³, $R_s \sim 10^6$ cm; for the $^3P_2$ anisotropic NSV, $\rho \sim (1-3) \times 10^{14}$ g/cm³, $R_P \sim 10^5$ cm. The angular momentums for these two types of NSV are:

$$J_{NSV}(^1S_0) \sim 1.7 \times 10^{28} \overline{n} \overline{\rho}_{12} R_6^3(^1S_0) \qquad (c.g.s.) \qquad (8)$$
$$J_{NSV}(^3P_2) \sim 4.7 \times 10^{27} \overline{n} (\overline{\rho}/\rho_{nuc}) R_5^3(^3P_2) \qquad (c.g.s.) \qquad (9)$$

It is well known that N vortices are energetically favorable be in the basic state of n=1 in the thermodynamically equilibrium of the superfluid vortices. However, the superfluid vortices in the young pulsars are in highly non-thermodynamics equilibrium state due to the significant fraction of the angular momentum of the rotational pre-supernova reserved in the turbulent classical vortices of the nascent neutron star during supernova explosion. Therefore, the initial quantum number $n_0$ of these quantized vortices (both $^1S_0$ and $^3P_2$ NSV) are very large after these phase transitions. For a typical NSV, $J_{NSV}(^3P_2) > 10^{-10} - 10^{-8} J(^3P_2)$, we get $\overline{n} > 10^4 - 10^6$

## 2.3  Heating of NS by the magnetic dipole radiation from the $^3P_2$ NSV

A magnetic dipole radiation is produced when a $^3P_2$ neutron Cooper pair with an abnormal magnetic momentum revolves around the axis of the neutron vortex (Peng, Huang and Huang,1980) and Huang, Lingenfelter, Peng and Huang,1982). The radiation frequency is the same as the rotational frequency of the superfluid neutron rotating around its vortex line, $\omega(r) = n\hbar(2m_n r^2)^{-1}$, ($\hbar$ is the Planck constant divided by $2\pi$, $m_n$ is the neutron mass, and $r$ is the radial distance from the vortex line). The radiation is mostly in the X-ray range and gets absorbed by the surrounding matter, which plays a role of heating the neutron star. The Heating rate is

$$W_{heat} = KQ(n)P_{SF}^{-1} \quad , \quad Q(n) = \overline{n^3}/\overline{n} \quad ,$$

$$K \approx 1 \times 10^{30}[(\frac{\Delta_n(^3P_2)}{0.05 MeV})^2 (\frac{B}{10^{13} gauss})^2]R_5^3(^3P_2) \quad (c.g.s) \tag{10}.$$

Here $B$ is strength of the magnetic field in the interior of the neutron star.

## 2.4 Phase Oscillation of $^3P_2$ NSV

**Normal neutron (Fermi) (turbulent) fluid $\Rightarrow$ $^3P_2$ NSV state**
The phase transition happens at time $t_0^{(P)}$ when $^3P_2$ NSV state start to emerge. The entire angular momentum deposited in the classical turbulent vortex is transferred to the angular momentum of the quantized vortices during this phase transition. Hence the initial quantum number of the $^3P_2$ NSV is also in $10^4 - 10^6$. Strong x-ray radiation is emitted by the rapidly rotating magnetic dipole of the $^3P_2$ Cooper pairs of neutrons revolving around the vortex line. The x-ray radiation is absorbed by the electrons in the normal Fermi gas on its way out, which heats the interior of the neutron star (Peng, Huang & Huang, 1980; Huang et al. 1982). If the heating rate by the x-ray exceeds the cooling rate by other physical processes, the temperature may increase. As an example, the modified Urca cooling rate is (Shapiro and Teukolsky, 1984)

$$L_\nu^{(Urca)} = 8.5 \times 10^{37} erg \cdot s^{-1} \cdot \frac{M}{M_{Sun}}(\frac{\rho_{nuc}}{\rho})^{1/3}(\frac{T}{6 \times 10^8 K})^8 \tag{11}$$

If the initial vortex quantum $n_0 > 10^3$ and $P_{SF}$ (the rotation period of $^3P_2$ superfluid around the axis of the neutron star) is *10 ms*, the heating power of the magnetic dipole radiation of $^3P_2$ SNV may exceeds $10^{38}$ erg/s. As long as the temperature from the heating mechanism goes slightly above the phase transition temperature, $T_\lambda(^3P_2(n))$, the $^3P_2$ Cooper pair will break up and the nascent $^3P_2$ NSV disappear immediately at the time $t = t_1^{(P)} > t_0^{(P)}$. The superfluid state suddenly returns to the normal Fermi gas state

$^3P_2$ **NSV state $\Rightarrow$ Normal neutron (Fermi) fluid**  (12)

At the same time, the quantized vortices of the $^3P_2$ NSV are transformed to the classical turbulent vortices with high angular momentum, hence the magnetic dipole radiation heating mechanism disappears. This imbalance in the thermodynamic equilibrium system then cools the interior of the neutron star again. The temperature of the normal neutron system would then shortly decrease to $T_\lambda(^3P_2(n))$, which is a little below the prior temperature just after the phase transition. The $^3P_2$ neutron (anisotropic) superfluid state appears again at the moment $t = t_2^{(P)} \geq t_1^{(P)}$, which shows a pattern of phase oscillation:

**$^3P_2$ NSV state $\Rightarrow$ Normal neutron (Fermi) fluid $\Rightarrow$ $^3P_2$ NSV state …** (13)

Emergence of the midway normal neutron Fermi turbulent fluid is transitory and the duration for its existence is very short: $\Delta t = (t_2^{(P)} - t_1^{(P)})$. The turbulent classical vortices get transformed into the quantized vortices of $^3P_2$ NSV again afterwards.
The phase oscillation repeats between these two states. It can be shown from Eq. (10) that, the quantum number n of the $^3P_2$ NSV in this case can be as high as $n > 10^3$, and then gradually decrease with a repeating phase oscillations. After n gets decreased to a level when heating rate by the magnetic dipole radiation of $^3P_2$ NSV no longer exceeds the cooling rate, the phase oscillation disappears and the $^3P_2$ NSV state no longer returns to the normal neutron Fermi fluid state.

## 3. Essence of Glitches

### 3.1 Kick-off of the first glitch:

It is believed that the interior superfluid rotates around the axis of NS faster than the crust. The difference in the two rotation rates can be maintained for a long time because the interaction between the superfluid neutrons and the plasma crust is very weak. This weak interaction originates from the interaction between the magnetic moment of the electrons either with the abnormal magnetic moment of normal neutrons in the core of the vortices or with the magnetic moment of $^3P_2$ neutron Cooper pair.

However, the scenario for the short-lived normal neutron fluid oscillated between the two successive phase transitions described in equation (13) is totally different. There is a strong coupling between the neutrons and the protons in the $^3P_2$ superfluid regions. This $^3P_2$ superfluid is non - superconductor which is in the outer and less dense regions than the proton - superconductor regions.

Due to this strong interaction, the short-lived rapid rotating normal neutron fluid in the deep interior will suddenly spin up the slow rotating plasma crust in the very short period of $\Delta t$. This explains the glitch phenomenon observed in the glitch pulsars in the following way:

The amplitude of the glitch $\Delta\Omega_0 / \Omega_0$, derived from the jumps of the crust rotation rate, is very small ($< 10^{-5}$). Hence the angular momentum transferred from the $^3P_2$ neutron superfluid with rapid rotation around the axis of the NS to the outer crust with slower rotation is much smaller than the original angular momentum of the outer crust. The amplitude of the glitch is determined by mainly following factors:
a) The difference between the rotating frequency of the primary $^3P_2$ superfluid prior to the phase transition and that of the crust.
b) The strength of the weak interaction of magnetic moment between the neutrons and the electrons.
c) The duration of the interaction, $\Delta t = (t_2^{(P)} - t_1^{(P)})$.

## 3.2 Repetition of glitches:

The duration for the reiteration of the phase transitions $\Delta t$ is very short, and the angular momentum deposited in the second classical turbulent vortices is still very large. Yet, the quantum number of the second $^3P_2$ NSV $n_2(^3P_2(n))$, is slightly smaller than that of the first $^3P_2$ NSV, i.e. $n_2(^3P_2(n)) < n_1(^3P_2(n))$. As long as the heating rate of the magnetic dipole radiation of $^3P_2$ SNV exceeds cooling rate, the following processes repeat again and again, with *n* number decreasing each time:

X-ray radiation comes from the magnetic dipole from the $^3P_2$ NSV → neutron star gets heated by the X-ray → the temperature in the neutron star increases and reaches the transition temperature $T_\lambda(^3P_2(n))$, → the superfluid returns to the normal neutron Fermi state through the sudden phase transition → glitch happens → the temperature of the neutron star decreases down to $T_\lambda(^3P_2(n))$ by cooling physical processes → the normal neutron Fermi fluid turns back to the $^3P_2$ SNV state through the phase transition.

The variation of the jump amplitude $\Delta\Omega_0/\Omega_0$ of the rotation rate for the successive glitches of the same pulsar is very small, since the angular momentum of outer crust is much smaller than that of the *$^3P_2$* superfluid. No obvious periodic law applies due to some stochastic factors in phase of the classical turbulent vortex neutron fluid, but the jump amplitude of glitches has a tendency to get decreased.

## 3.3 Disappearing of Glitches

The interval of successive glitches is decided by the competition of the heating rate of the x-ray by the *$^3P_2$ NSV* against the cooling process rate. As described in the previous section, the quantum number, $n_i(^3P_2(n))$ of the i$^{th}$ *$^3P_2$ NSV* is smaller than that of the *(i-1)*$^{th}$ *$^3P_2$ NSV*, so the heating rate of the x-ray by the *$^3P_2$ NSV* is getting lower each time. The phase oscillation of neutron system will finally stop when the heating rate of the x-ray by the *$^3P_2$ NSV* is lower than the cooling rate. At this point, the *$^3P_2$ NSV* state will no longer return to the normal state, hence the glitches appears. This results agrees with the observational facts that no glitch have been detected in the old pulsars with period *P > 0.7s* (Lyne, 2000)